\documentclass[11pt]{article}

\usepackage{amssymb,amsmath}

\makeatletter
\@addtoreset{equation}{section}

\makeatother

\newtheorem{Thm}{Theorem}
\newtheorem{Lem}{Lemma}

\DeclareRobustCommand{\qed}{%
  \ifmmode 
  \else \leavevmode\unskip\penalty9999 \hbox{}\nobreak\hfill
  \fi
  \quad\hbox{\qedsymbol}}
\newcommand{\qedsymbol}{\leavevmode
  \hbox to.77778em{\hfil\vrule \vbox to.675em{\hrule width.6em\vfil\hrule}%
  \vrule\hfil}}

\newenvironment{proof}[1][Proof]
{\par \normalfont \trivlist \item[\hskip\labelsep\itshape #1.]\ignorespaces}
{\qed\endtrivlist}

\newcommand{\C}{\mathbb C}
\newcommand{\R}{\mathbb R}
\newcommand{\N}{\mathbb N}
\newcommand{\Z}{\mathbb Z}
\renewcommand{\H}{\mathcal H}
\newcommand{\B}{\mathcal B}
\newcommand{\bra}[1]{\mbox{$\left\langle #1 \right|$}}
\newcommand{\ket}[1]{\mbox{$\left| #1 \right\rangle$}}

\begin{document}
\title{Quantum Database Searching by a Single Query}
\author{
 \textsc{Dong Pyo Chi}
        \thanks{Department of Mathematics,
                Seoul National University, Seoul 151-742, Korea.}
        \thanks{E-mail address: \texttt{dpchi@math.snu.ac.kr}}
 \ and
 \textsc{Jinsoo Kim}
        \footnotemark[1]
        \thanks{E-mail address: \texttt{jskim@math.snu.ac.kr}}
}
\date{15 July 1997}
\maketitle

\begin{abstract}
In this paper we give a quantum mechanical algorithm that can
search a database by a single query,
when the number of solutions is more than a quarter.
It utilizes modified Grover operator of arbitrary phase.
\end{abstract}

\section{Introduction}

For $N \in \N$, let $\Z_N=\{ 0, 1, \dots, N-1 \}$ denote
the additive cyclic group of order $N$
and consider an arbitrary function $F: \Z_N \rightarrow \Z_2$,
The database searching problem is to find some $i \in \Z_N$
such that $F(i) = 1$ under the assumption that such an $i$ exists.
We assume that the structure of $F$ is unknown so that
it is not possible to obtain a knowledge about $F$ without evaluating it
on $\Z_N$.

Let $t= \left| \{ i \in \Z_N \, : \, F(i) = 1 \} \right|$.
There is a quantum mechanical algorithm to solve this problem
in expected time of order $O(\sqrt{N/t})$,
which is optimal up to a multiplicative constant
\cite{Grover96final, BBHT96, BBBV95}.
Especially when $t=N/4$ is known,
the original Grover algorithm in \cite{Grover96final}
can search a solution only by a single query \cite{BBHT96}.
It uses the $\pi$-phase, i.e.,
marking the states by multiplying $e^{\pi\imath}=-1$.
When $t=N/2$,
by changing this phase to $\pi/2$, that is,
by marking the states by multiplying $e^{\pi\imath/2} = \imath$
and modifying the corresponding diffusion transform according to this phase,
the solution can be found with certainty after a single iteration
\cite{BH97}.

In this paper, we generalize Grover algorithm for arbitrary phase.
When $t \ge N/4$,
we give generalized conditional phase and diffusion transform
depending on $t$, and then formulate a quantum mechanical algorithm
that solves the database searching problem in a single query.

\section{Grover Operator of Arbitrary Phase} \label{sec:grover}

Let
$\B_{N} = \{ \ket{a} \}_{a\in \Z_2^n}
= \{ \ket{e_1}, \ket{e_2}, \dots, \ket{e_N} \}$ be the standard basis
of an $n$-qubit quantum register with $N=2^n$
and $\H_{N}$ be the corresponding Hilbert space,
which represents the state vectors of a quantum system.
Let $\H_m$ be an $m$-dimensional subspace of $\H_N$
spanned by a basis
$\B_m=\{ \ket{e_{i_1}}, \ket{e_{i_2}},\dots,\allowbreak \ket{e_{i_m}} \}$.
Let $l$ be a positive integer such that $1 \le l \le m$.

For $\gamma \in \R$,
the {\em conditional $\gamma$-phase transform\/} on a subspace $\H_m$,
$\mathbf S_{F,\gamma}^{\H_m}: \H_m \rightarrow \H_m$ is defined by
\[
  \mathbf S_{F,\gamma}^{\H_m} \ket{e_{i_k}}
  = (e^{\imath\gamma})^{F(e_{i_k})} \ket{e_{i_k}} \, ,
\]
for $k=1,2,\dots,m$, where $\imath=\sqrt{-1}$.
Let $\mathbf S_{l,\gamma}^{\H_m}$ denote
$\mathbf S_{F_l,\gamma}^{\H_m}$,
where $F_l(e_{i_k})= \delta_{i_k i_l}$.

Let $\mathbf W_l^{\H_m}$ be any unitary transformation on $\H_m$ satisfying
\[
  \mathbf W_l^{\H_m} \ket{e_{i_l}}
  = \frac{1}{\sqrt{m}} \sum_{k=1}^{m} \ket{e_{i_k}} \, .
\]
For example, when $m$ is a power of 2, in a suitably arranged basis
we may set $\mathbf W_1^{\H_m}$ to be the Walsh-Hadamard transform.
When $m$ is not a power of 2, the approximate Fourier transform in
\cite{Kitaev95} can be used.

For $\beta \in \R$,
the {\em $\beta$-phase diffusion transform\/} on a subspace $\H_m$,
$\mathbf D_{\beta}^{\H_m}:\H_N \rightarrow \H_N$ is defined by
\[
 \mathbf D_{\beta,ij}^{\H_m}
 =\bra{e_i} \mathbf D_{\beta}^{\H_m} \ket{e_j}
 =\begin{cases}
  \frac{e^{\imath\beta}-1}{m}
    &\text{when $\ket{e_i},\ket{e_j}\in \B_m$ and $i\ne j$},\\
  1+\frac{e^{\imath\beta}-1}{m}
    &\text{when $\ket{e_i},\ket{e_j}\in \B_m$ and $i=j$},\\
  \delta_{ij}         &\text{otherwise} \, .
  \end{cases}
\]
If we rearrange the basis $\B_N$ so that
$\B_m = \{ \ket{e_1}, \ket{e_2}, \dots, \ket{e_m} \}$
and represent $\mathbf D_{\beta}^{\H_m}$ by its matrix
$\mathbf D_{\beta}^{\H_m} = ( \mathbf D^{\H_m}_{\beta,ij} )$,
then we have
\begin{align*}
  \allowdisplaybreaks
  \mathbf D_{\beta}^{\H_m}
  & = \begin{bmatrix}
      \mathbf W_l^{\H_m} \mathbf S_{l,\beta}^{\H_m}
      \mathbf W_l^{\H_m \dagger} & 0 \\
      0 & I
      \end{bmatrix} \\ 
  & = \begin{bmatrix}
      I + (e^{\imath\beta}-1) P^{\H_m} & 0 \\
      0 & I
      \end{bmatrix}  ,
\end{align*}
where $P^{\H_m} =(P_{ij}^{\H_m})$ is a projection matrix in $\H_m$
with $P_{ij}^{\H_m} = \frac{1}{m}$.
Note that $\mathbf S_{F,\gamma}^{\H_m}$ and $\mathbf D_{\beta}^{\H_m}$
are unitary.

Let $\mathbf G_{F,\beta,\gamma}^{\H_m}:\H_N \rightarrow \H_N$ be
the {\em Grover operator\/} of $(\beta,\gamma)$-phase in $\H_m$ defined by
\[
 \mathbf G_{F,\beta,\gamma}^{\H_m}
 = \mathbf D_{\beta}^{\H_m} \mathbf S_{F,\gamma}^{\H_m} \, .
\]
When $\beta=\gamma$ set
$\mathbf G_{F,\gamma}^{\H_m} = \mathbf G_{F,\gamma,\gamma}^{\H_m}$.
For simplicity, we shall assume that $m = N$
and omit the superscript $\H_m$.

Let
\begin{align*}
  A &= \{ e_j \in \Z_2^n | F(e_j)=1 \} \, , \\
  B &= \{ e_j \in \Z_2^n | F(e_j)=0 \} \, ,
\end{align*}
and let $t=|A|$.
For $k,l \in \C$ such that $t |k|^2 + (N-t) |l|^2 = 1$,
define
\[
 \ket{\psi(k,l)} = \sum_{e_j\in A}k\ket{e_j}+\sum_{e_j\in B}l\ket{e_j} \, .
\]

\begin{Lem}
For $\beta, \gamma \in \R$, let
\[
 \ket{\psi(k_j,l_j)} = \mathbf G_{F,\beta,\gamma}^j \ket{\psi(k_0,l_0)} \, .
\]
Then after applying $j+1$ Grover operator of $(\beta,\gamma)$-phase,
we have
\begin{equation} \label{eq:amp}
\left\{
\begin{aligned}
 k_{j+1} &=  \frac{(e^{\imath\beta}-1)t+N}{N}e^{\imath\gamma} k_j
            +\frac{(e^{\imath\beta}-1)(N-t)}{N} l_j  \, , \\
 l_{j+1} &=  \frac{(e^{\imath\beta}-1)t}{N}e^{\imath\gamma} k_j
            +\frac{(e^{\imath\beta}-1)(N-t)+N}{N} l_j \, .
\end{aligned} \right.
\end{equation}
\end{Lem}

\begin{Thm} \label{thm:amp}
Assume that $k_0 = l_0$ and $\beta, \gamma \in [0, 2 \pi]$.
Then $l_1 = 0$ if and only if
$\frac{N}{4} \le t \le N$ and
$\beta= \gamma = \cos^{-1} \left(1 - \frac{N}{2t} \right)$.
In this case, $k_1 = (e^{\imath\gamma}-1) k_0$.
\end{Thm}

\begin{proof}
By~\eqref{eq:amp}, we get
\begin{equation} \label{eq:amp1}
\left\{
\begin{aligned}
 k_1 &= \left[ (e^{\imath\beta}-1)(e^{\imath\gamma}-1) \frac{t}{N}
        + e^{\imath\gamma} + e^{\imath\beta} - 1 \right] k_0 \, , \\
 l_1 &= \left[ (e^{\imath\beta}-1)(e^{\imath\gamma}-1) \frac{t}{N}
        + e^{\imath\beta} \right] l_0 \, .
\end{aligned} \right.
\end{equation}
By considering the imaginary part of $e^{-\imath\beta} \frac{l_1}{l_0}$,
the equation
\[
 \frac{t}{N} \left\{
 (1-\cos\beta) \sin\gamma + (\cos\gamma-1)\sin\beta
 \right\} =0
\]
is equivalent to
\begin{equation} \label{eq:im}
 \frac{1-\cos\beta}{\sin\beta} = \frac{1-\cos\gamma}{\sin\gamma} \, .
\end{equation}
Considering the real part of $e^{-\imath\beta} \frac{l_1}{l_0}$,
from~\eqref{eq:im} it follows that the equation
\[
 \frac{t}{N} \left\{
 (1-\cos\beta) (1-\cos\gamma) + \sin\beta \sin\gamma
 \right\} - 1 =0
\]
is equivalent to
\[
 \cos\beta = \cos\gamma = 1 - \frac{N}{2t} \, .
\]
Again by~\eqref{eq:im}, we get $\beta = \gamma$
and $t \ge \frac{N}{4}$.
Furthermore, by~\eqref{eq:amp1} we obtain $k_1 = (e^{\imath\gamma}-1) k_0$.
This completes the proof.
\end{proof}

For the case of $\pi$-phase in \cite{Grover96final},
$ - \mathbf D_{\pi} = - I + 2 P$ is an inversion about average operation
and we have
\[
 - \mathbf D_{\pi} \ket{\psi(k,l)}
 = \ket{\psi(-\frac{N-2t}{N}k + \frac{2(N-t)}{N}l,
              \frac{N-2t}{N}l + \frac{2t}{N}k)} \, .
\]
In this case, there is an explicit closed-form formula for $k_j$ and $l_j$;
\[
\left\{
\begin{aligned}
 k_j &= (-1)^j \frac{1}{\sqrt{t}}   \sin \left( (2j+1)\theta \right) \, , \\
 l_j &= (-1)^j \frac{1}{\sqrt{N-t}} \cos \left( (2j+1)\theta \right) \, ,
\end{aligned} \right.
\]
for $j=0,1,\dots$, where the angle $\theta$ is defined so that
$\sin^2 \theta = \frac{t}{N}$ \cite{BBHT96}.
Especially when $t = N/4$, we have $l_1 = 0$.

Grover operator of $\frac{\pi}{2}$-phase was used in \cite{BH97}.
Since
\[
 \mathbf D_{\frac{\pi}{2}} \ket{\psi(k,l)}
 = \ket{\psi(\frac{(\imath-1)t+N}{N}k + \frac{(\imath-1)(N-t)}{N}l,
             \frac{(\imath-1)(N-t)+N}{N}l + \frac{(\imath-1)t}{N}k)} \, ,
\]
when $t=\frac{N}{2}$, we have
\[
 \mathbf G_{F,\frac{\pi}{2}} \ket{\psi(k,k)}
 = \ket{\psi( (\imath-1)k,0 )} \, .
\]

By Theorem~\ref{thm:amp},
When $t \in [ N/4, N]$, we have
\[
 \mathbf G_{F,\gamma} \ket{\psi(k_0,k_0)}
 = \ket{\psi((e^{\imath\gamma}-1) k_0, 0)} \, ,
\]
where the phase $\gamma$ is defined by
$\gamma = \cos^{-1} \left(1 - \frac{N}{2t} \right)$.



\begin{thebibliography}{1}

\bibitem{BBBV95}
C.~H. Bennett, E.~Bernstein, G.~Brassard and U.~Vazirani,
\emph{Strengths and weaknesses of quantum computing},
Los Alamos e-print quant-ph/9701001, 1997.

\bibitem{BBHT96}
M.~Boyer, G.~Brassard, P.~H{\o}yer and A.~Tapp,
\emph{Tight bounds on quantum searching},
Proceedings of the Fourth Workshop on Physics and Computation,
New England Complex Systems Institute, 1996, pp.~36--43;
Los Alamos e-print quant-ph/9605034, 1996.

\bibitem{BH97}
G.~Brassard and P.~H{\o}yer,
\emph{An exact quantum polynomial-time algorithm for {S}imon's problem},
Proceedings of the Fifth Israeli Symposium on Theory of Computing and Systems,
1997, (to appear);
Los Alamos e-print quant-ph/9704027, 1997.

\bibitem{Grover96final}
L.~K. Grover,
\emph{Quantum mechanics helps in searching for a needle in a haystack},
Phys. Rev. Lett., (to appear);
Los Alamos e-print quant-ph/9605043, 1996.

\bibitem{Kitaev95}
A.~Yu. Kitaev,
\emph{Quantum measurements and the abelian stabilizer problem},
Los Alamos e-print quant-ph/9511026, 1995.

\end{thebibliography}
\end{document}